\documentclass[prl, reprint]{revtex4-1}

\usepackage{graphicx}
\usepackage{indentfirst}
\usepackage{latexsym}
\usepackage{multirow}
\usepackage{tabls}
\usepackage{color}
\usepackage{subfigure}
\usepackage{comment}

\usepackage{amsfonts}
\usepackage{amsmath}
\usepackage{bm}
\usepackage{mathrsfs}

\def\b{ \beta }

\def\b0{ {\bf 0} }

\def\veceta{ {\vec{\eta} }  }
\def\vecq{ {\vec{q} } }
\def\vecp{ {\vec{p} } }
\def\vecpi{ {\vec{\pi} } }
\def\vecv{ {\vec{v} } }
\def\vecx{ {\vec{x} } }

\def\lsim{\mathrel{\rlap{\lower3pt\hbox{\hskip1pt$\sim$}}
    \raise1pt\hbox{$<$}}}                
\def\gsim{\mathrel{\rlap{\lower3pt\hbox{\hskip1pt$\sim$}}
    \raise1pt\hbox{$>$}}}         

\def\coordeq{ \, \mathrel{ \rlap{\hbox{\hskip-2.5pt$=$} }
    \raise4pt\hbox{$\cdot$}} \, }                

\begin{document}

\title{The classical mechanics of non-conservative systems}

\author{Chad R. Galley}
\email{crgalley@tapir.caltech.edu}

\affiliation{Jet Propulsion Laboratory, California Institute of Technology, Pasadena, CA, 91109}
\affiliation{Theoretical Astrophysics, California Institute of Technology, Pasadena, CA, 91125}

\begin{abstract}
	Hamilton's principle of stationary action lies at the foundation of theoretical physics and is applied in many other disciplines from pure mathematics to economics. Despite its utility, Hamilton's principle has a subtle pitfall that often goes unnoticed in physics:  it is formulated as a boundary value problem in time but is used to derive equations of motion that are solved with initial data. This subtlety can have undesirable effects. I present a formulation of Hamilton's principle that is compatible with initial value problems. Remarkably, this leads to a natural formulation for the Lagrangian and Hamiltonian dynamics of generic non-conservative systems, thereby filling a long-standing gap in classical mechanics. Thus dissipative effects, for example, can be studied with new tools that may have application in a variety of disciplines. The new formalism is demonstrated by two examples of non-conservative systems: an object moving in a fluid with viscous drag forces and a harmonic oscillator coupled to a dissipative environment.
\end{abstract}

\maketitle

Hamilton's principle of stationary action \cite{Goldstein} is a cornerstone of physics and is the primary, formulaic way to derive equations of motion for many systems of varying degrees of complexity -- from the simple harmonic oscillator to supersymmetric gauge quantum field theories. 
Hamilton's principle relies on a Lagrangian or Hamiltonian formulation of a system, which account for conservative dynamics but cannot describe generic
non-conservative interactions. For simple dissipation forces local in time and linear in the velocities, one may use Rayleigh's dissipation function \cite{Goldstein}. However, this function is not sufficiently comprehensive to describe systems with more general dissipative features like history-dependence, nonlocality, and nonlinearity that can arise in open systems.

The dynamical evolution and final configuration of non-conservative systems must be determined from initial conditions. 
However, it seems under-appreciated that while initial data may be used to solve equations of motion derived from Hamilton's principle, the latter is formulated with {\it boundary} conditions in time, not initial conditions. This observation may seem innocuous, and it usually is, except that this subtlety may manifest undesirable features. Remarkably, resolving this subtlety opens the door to proper Lagrangian and Hamiltonian formulations of {\it generic non-conservative} systems.

~\\
\noindent {\it An illustrative example}. To demonstrate the shortcoming of Hamilton's principle, consider a harmonic oscillator with amplitude $q(t)$, mass $m$, and frequency $\omega$ coupled with strength $\lambda$ to another harmonic oscillator with amplitude $Q (t)$, mass $M$, and frequency $\Omega$. The action for this system is 
\begin{align}
	S [ q, Q ] = {} & \int_{t_i}^{t_f} \!\!\! dt \bigg\{ \frac{m}{2} \big( \dot{q}^2 - \omega^2 q^2 \big) + \lambda q Q \nonumber \\
		& {\hskip0.35in} + \frac{ M}{2} \big( \dot{Q}^2 - \Omega^2 Q^2 \big)  \bigg\} .
\label{eq:action1}
\end{align}
The total system conserves energy and is Hamiltonian but $q(t)$ \emph{itself} is open to exchange energy with $Q$ and should thus be non-conservative. For a large number of $Q$ oscillators the open (sub)system dynamics for $q$ ought to be dissipative. 

Let us account for the effect of the $Q$ oscillator on $q(t)$ by finding solutions only to the equations of motion for $Q$ and inserting them back into (\ref{eq:action1}), which is called {\it integrating out}. The resulting action, 
\begin{align}
	 S_{\rm eff} [ q ] = {} &  \int_{t_i}^{t_f} \!\!\! dt \bigg\{ \frac{ m}{2} \big( \dot{q}^2 - \omega^2 q^2 \big) + \lambda q  Q^{(h)}(t)  \nonumber \\
		& {\hskip0.35in} + \frac{ \lambda^2 }{ 2 M } \int_{t_i}^{t_f} \!\!\! dt' \, q(t)  G_{\rm ret} (t-t') q(t') \bigg\} ,
\label{eq:effaction1}
\end{align}
is the {\it effective action} for $q(t)$, though it is sometimes called a Fokker action \cite{WheelerFeynman}. $Q^{(h)}(t)$ is a homogeneous solution (from initial data) and $G_{\rm ret} (t-t')$ is the retarded Green function for the $Q$ oscillator. 

The last term in (\ref{eq:effaction1}) involves two time integrals and the product $q(t) q(t')$. The latter is symmetric in $t \leftrightarrow t'$ and couples only to the time-symmetric part of the retarded Green function. Hence, the last term in (\ref{eq:effaction1}) equals
\begin{align}
	  \frac{ \lambda^2 }{ 2 M } \int_{t_i}^{t_f} \!\!\! dt dt' \, q(t) \bigg[ \frac{ G_{\rm ret} (t-t') + G_{\rm adv} (t-t') }{ 2 } \bigg] q(t') 
\label{eq:effaction1a}
\end{align}
when using the identity $G_{\rm ret} (t' - t) = G_{\rm adv}(t-t')$. Applying Hamilton's principle to the effective action (\ref{eq:effaction1}) yields the equation of motion for $q(t)$
\begin{align}
	m \ddot{q} + m \omega^2 q = {} & \lambda Q^{(h)} (t)  + \frac{ \lambda^2}{2M} \int_{t_i}^{t_f} \!\!\! dt' \nonumber \\
		& \times \bigg[ G_{\rm ret} (t-t') + G_{\rm adv} (t-t') \bigg] q(t') .
\label{eq:qeom1}
\end{align}
There are a couple of key points regarding (\ref{eq:qeom1}). First, the second term on the right side depends on the advanced Green function implying that solutions to (\ref{eq:qeom1}) do not evolve causally nor are specified by initial data alone. 
Second, the kernel of the integral in (\ref{eq:qeom1}) is symmetric in time, which means that the integral describes {\it conservative} interactions  between $q$ and $Q$. Consequently, (\ref{eq:qeom1}) does not account for dissipation, a time-asymmetric process, that should be present  when there are $N>>1$ of the $Q$ oscillators. 

\begin{figure}
	\includegraphics[width=0.94\columnwidth]{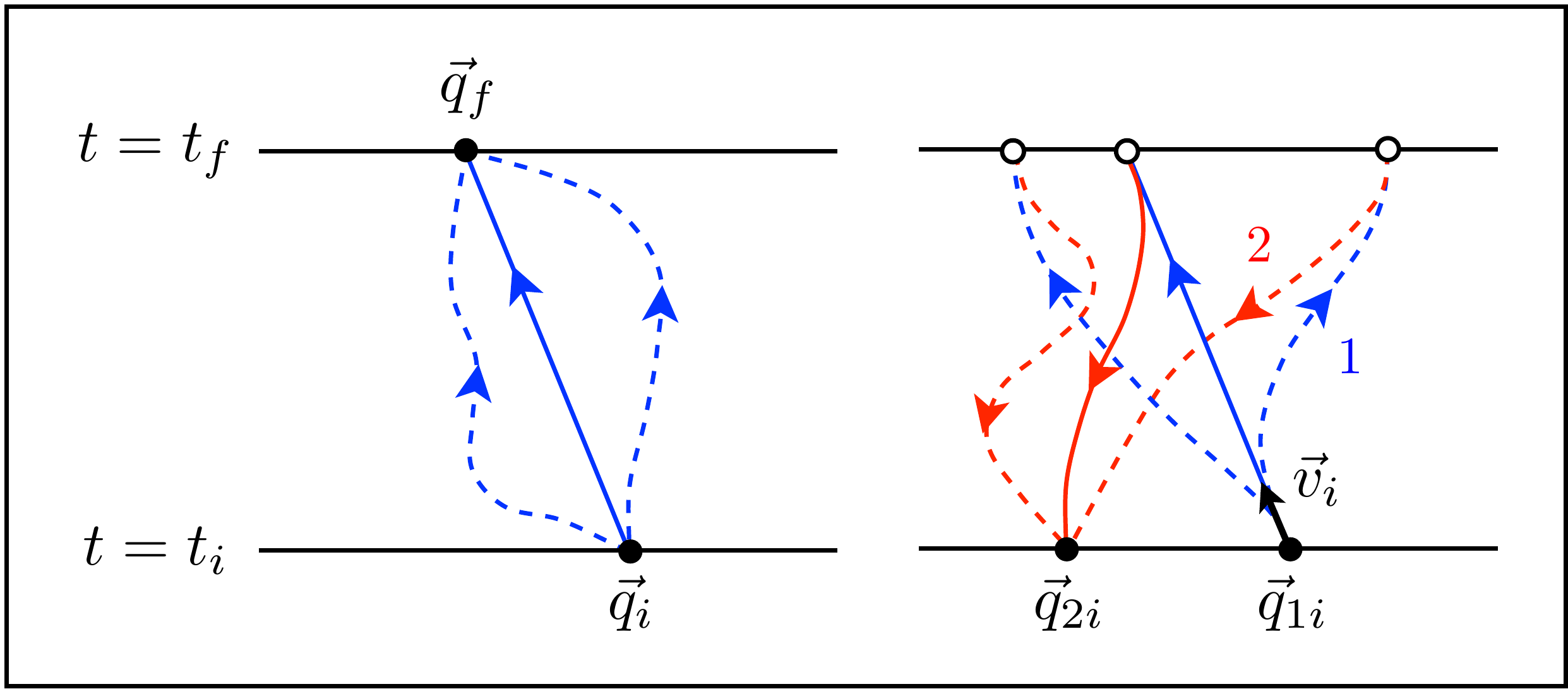}
	\caption{{\bf Left}: A cartoon of Hamilton's principle. Dashed lines denote the virtual displacements and the solid line the stationary path.
	{\bf Right}: A cartoon of Hamilton's principle compatible with initial data (i.e., the final state is not fixed). In both cartoons, the arrows on the paths indicate the integration direction for the line integral of the Lagrangian.}
	\label{fig:cartoon}
\end{figure}

These undesirable features can be traced back to the very formulation of Hamilton's principle, which solves the problem: ``Find the path $\vecq(t)$ passing through the given values $\vecq_i$ at $t=t_i$ and $\vecq_f$ at $t=t_f$ that makes the action stationary'' (see left cartoon in Fig.~\ref{fig:cartoon}). Stated in this way, it is clear that Hamilton's principle is appropriate for systems satisfying {\it boundary} conditions in time, not {\it initial} conditions. 
According to Sturm-Liouville theory \cite{Arfken}, the {\it time-symmetric} integration kernel in (\ref{eq:qeom1}), which is a Green function itself, satisfies {\it boundary} conditions in time. 
Likewise, boundary conditions in time imply the corresponding Green function is time-symmetric.
This example indicates an intimate connection in the variational calculus between boundary (initial) conditions and conservative (non-conservative) dynamics. 

In the remainder I formulate Hamilton's principle with {\it initial} conditions for {\it general} systems, report some consequences, and present some examples.

~\\
\noindent {\it Hamilton's principle with initial data}. 
A hint for how to proceed comes from the previous example.
The advanced Green function in (\ref{eq:effaction1a}) and (\ref{eq:qeom1}) appears because the factor $q(t) q(t')$ couples only to the time-symmetric part of the retarded Green function. 
``Breaking'' the symmetry by introducing two sets of variables, say $q_1$ and $q_2$, implies that $q_1(t) q_2(t')$ will couple to the full retarded Green function, not just its time-symmetric part. Varying with respect to only $q_1$ gives the correct force provided one sets $q_2 = q_1$ after the variation 
\footnote{Doubled variables appear in specific applications as early as \cite{Staruszkiewicz:1970} and more recently in \cite{JaranowskiSchafer:PRD55, KonigsdorfferFayeSchafer:PRD68}.
However, this paper constructs a general framework relating the doubled variables, Hamilton's principle, initial conditions, and non-conservative dynamics for general dynamical systems.
}.
This procedure is formalized and developed for general systems below.

Let $\vecq \equiv \{q_i \}_{i=1}^N$ and $\dot{\vecq} \equiv \{ \dot{q}_i \}_{i=1}^N$ be a set of $N$ generalized coordinates and velocities of a general dynamical system. 
Formally, double both sets of quantities, $\vecq \to (\vecq_1, \vecq_2)$ and $\dot{\vecq} \to (\dot{\vecq}_1, \dot{\vecq}_2)$. Parameterize both coordinate paths as $\vecq_{1,2}(t,\epsilon) = \vecq_{1,2}(t,0) + \epsilon \veceta_{1,2}(t)$ where $\vecq_{1,2}(t,0)$ are the coordinates of the two stationary paths, $\epsilon \ll 1$, and $\veceta_{1,2}(t)$ are arbitrary virtual displacements. To ensure that enough conditions are given for varying the action we require that: 1) $\veceta_{1,2}(t_i) = 0$ and 2) $\vecq_1 (t_f, \epsilon) = \vecq_2(t_f, \epsilon)$ and $\dot{\vecq}_1 (t_f, \epsilon) = \dot{\vecq}_2 (t_f, \epsilon)$ for all $\epsilon$ (the {\it equality condition}). The equality condition does not fix either value at the final time since the values they equal are not specified. After all variations are performed, both paths are set equal to each other and identified with the physical one, $\vecq(t)$ (the {\it physical limit}). See the right cartoon in Fig.~\ref{fig:cartoon}.

The action functional of $\vecq_1$ and $\vecq_2$ is defined here as the total line integral of the Lagrangian along both paths plus the line integral of a function $K$ (discussed below) that depends on both paths $\{\vecq_a \}_{a=1}^2$ and cannot generally be written as the difference of two potentials, 
\begin{align}
	{\hskip-0.05in} S [ \vecq_a ] \equiv {} & \!\! \int_{t_i}^{t_f} \!\!\!\! dt \, L(\vecq_1, \dot{\vecq}_1) + \!\! \int_{t_f}^{t_i} \!\!\!\! dt \, L(\vecq_2, \dot{\vecq}_2)   + \!\! \int_{t_i}^{t_f} \!\!\!\! dt \, K (\vecq_a, \dot{\vecq}_a, t ) \nonumber \\
		= {} & \!\! \int_{t_i}^{t_f} \!\!\!\!  dt \big[ L(\vecq_1 , \dot{\vecq}_1 ) - L (\vecq_2, \dot{\vecq}_2) + K (\vecq_a, \dot{\vecq}_a, t ) \big]   .
\label{eq:action3}
\end{align}
This action defines a new Lagrangian 
\begin{align}
	\Lambda (\vecq_a, \dot{\vecq}_a ) \equiv L(\vecq_1, \dot{\vecq}_1) - L (\vecq_2, \dot{\vecq}_2) + K (\vecq_a, \dot{\vecq}_a , t).
\label{eq:lagrangian1}
\end{align}
If $K$ could be written as the difference of two potentials, $V(\vecq_1) - V(\vecq_2)$, then it could be absorbed into the difference of the Lagrangians in (\ref{eq:action3}), leaving $K$ zero \footnote{
The same is true in the general case if $K = V_1(\vecq_1, \dot{\vecq}_1) - V_2 (\vecq_2, \dot{\vecq}_2)$ for $V_1 \ne V_2$. The resulting equations of motion are the same as if $(V_1 + V_2)/2$ is absorbed into each Lagrangian.
}. Thus, a non-zero $K$ describes generalized forces that are {\it not} derivable from a potential (i.e., non-conservative forces) and couples the two paths with each other. 

It is convenient, but not necessary, to make a change of variables to $\vecq_+ = (\vecq_1+ \vecq_2)/2$ and $\vecq_- = \vecq_1 - \vecq_2$ because $\vecq_- \to 0$ and $\vecq_+ \to \vecq$ in the physical limit. The conjugate momenta in the ``$\pm$'' variables, regarded as functions of the ``$\pm$'' coordinates and velocities, are found to be $\vecpi_{\pm } = \partial \Lambda / \partial \dot{\vecq}_{\mp}$, and the paths are parameterized as $\vecq_\pm (t, \epsilon) = \vecq_\pm (t, 0) + \epsilon \veceta_\pm(t)$. The new action (\ref{eq:action3}) is stationary under these variations if $0 = [ d S [ \vecq_\pm] / d\epsilon ]_{\epsilon=0}$ for all $\veceta_\pm$, or
\begin{align}
	0 = {} &  \int_{t_i}^{t_f} \!\!\! dt \bigg\{ \veceta_+ \cdot \bigg[ \frac{ \partial \Lambda }{ \partial \vecq_+} - \frac{ d \vecpi_-}{dt} \bigg]_{0}  + \veceta_- \cdot \bigg[ \frac{ \partial \Lambda }{ \partial \vecq_-} - \frac{ d \vecpi_+}{dt} \bigg]_{0} \bigg\} \nonumber \\
	& + \bigg[ \veceta_+(t) \cdot \vecpi_-(t) + \veceta_-(t) \cdot \vecpi_+(t) \bigg]_{t=t_i}^{t_f}
\label{eq:variation2}
\end{align}
where the subscript $0$ denotes evaluation at $\epsilon=0$ and $\veceta_+ \cdot \vecpi_- = \sum_{i=1}^N \eta_{+i} \pi_{-i}$, etc.

The equality condition requires $\vecq_1(t_f, \epsilon) = \vecq_2 (t_f, \epsilon)$ and $\dot{\vecq}_1(t_f, \epsilon) = \dot{\vecq}_2 (t_f, \epsilon)$ so that $\veceta_- (t_f) = 0$ and $\vecpi_- (t_f) = 0$. With $\veceta_\pm (t_i ) = 0$ it follows that the boundary terms in (\ref{eq:variation2}) all vanish. Thus, (\ref{eq:variation2}) is satisfied for any $\vec{\eta}_\pm(t)$ provided that the two variables $\vecq_\pm (t)$ solve 
\begin{align}
	\frac{ d \vecpi_\mp }{ dt} = \frac{ \partial \Lambda }{ \partial \vecq_\pm}  .
\label{eq:EL2}
\end{align}
Of course, one could have used the $\vecq_{1,2}$ coordinates instead to find $d\vecpi_{1,2} / dt = \partial \Lambda / \partial \vecq_{1,2}$ with $\vecpi_{1,2} = (-1)^{1,2}\partial \Lambda / \partial \dot{\vecq}_{1,2}$ regarded as functions of $\vecq_{1,2}$ and $\dot{\vecq}_{1,2}$.

In the physical limit (``p.l.''), only the $d\vecpi_+/dt = \partial \Lambda / \partial \vec{q}_-$ equation in (\ref{eq:EL2}) survives, yielding
\begin{align}
	\frac{ d \vecpi ( \vecq, \dot{\vecq} \, )  }{ dt } = \bigg[ \frac{ \partial \Lambda }{ \partial \vecq_- } \bigg] _{ {\rm p.l.} } = \frac{ \partial L }{ \partial \vecq } + \bigg[ \frac{ \partial K }{ \partial \vecq_- } \bigg] _{ {\rm p.l.} } ~, 
\label{eq:newEL1}
\end{align}	
where the conjugate momenta are 
\begin{align}
	 \vecpi ( \vecq, \dot{\vecq} \, ) = \bigg[ \frac{ \partial \Lambda }{ \partial \dot{\vecq}_- } \bigg] _{ {\rm p.l.} }  =  \frac{ \partial L }{ \partial \dot{\vecq} } + \bigg[ \frac{ \partial K }{ \partial \dot{\vecq}_- } \bigg] _{ {\rm p.l.} } ~.
\label{eq:newp1}
\end{align}
When $K=0$ the generalized forces are derived from potentials and one recovers the usual Euler-Lagrange equations. A non-zero $K$ can be regarded as a ``non-conservative potential.''

In the physical limit, only the Euler-Lagrange equation for the ``$+$'' variable survives. Hence, expanding the action in powers of $\vecq_-$, the equations of motion in (\ref{eq:newEL1}) and (\ref{eq:newp1}) also follow from the variational principle
\begin{align}
	0 = \bigg[ \frac{ \delta S [ \vecq_\pm] }{ \delta \vecq_- (t) } \bigg]_{ {\rm p.l.} } ~.
\label{eq:shortcut1}
\end{align}
Only terms in the new action (\ref{eq:action3}) that are perturbatively {\it linear} in $\vecq_-$ contribute to physical forces. 

A new Hamiltonian $A$ is defined by Legendre transforming the new Lagrangian with respect to the usual conjugate momenta for each path, $\vecp_1$ and $\vecp_2$, \footnote{Using Legendre transforms with respect to $\vecp_a$ or $\vecpi_a$ leads to different but related Hamiltonian formulations that will be detailed elsewhere.}
\begin{align}
	A ( \vecq_{1,2}, \vecp_{1,2} ) & \equiv \vecp_1 \!\cdot\! \dot{\vecq}_1 - \vecp_2 \!\cdot\! \dot{\vecq}_2 - \Lambda ( \vecq_{1,2}, \dot{\vecq}_{1,2} )
\label{eq:hamiltonian1}  \\
		& =  H ( \vecq_1, \vecp_1 ) - H(\vecq_2, \vecp_2)  - K ( \vecq_{1,2} , \dot{\vecq}_{1,2} , t) \nonumber
\end{align}
where $\dot{\vecq}_{1}$ and $\dot{\vecq}_2$ are now functions of their respective coordinates and momenta.
Writing (\ref{eq:hamiltonian1}) in the ``$\pm$'' variables gives
\begin{align}
	A ( \vecq_\pm, \vecp_\pm ) = \vecp_+ \!\cdot\! \dot{\vecq}_- +  \vecp_- \!\cdot\! \dot{\vecq}_+ - \Lambda ( \vecq_\pm, \dot{\vecq}_\pm )  .
\label{eq:hamiltonian1pm}
\end{align}
Both (\ref{eq:hamiltonian1}) and (\ref{eq:hamiltonian1pm}) can be written as
\begin{align}
	A(\vecq_a , \vecp_a ) = {} & \vecp_a \dot{\vecq}^{\,a} - \Lambda (\vecq_a, \dot{\vecq}_a )
\label{eq:hamiltonian2}  
\end{align}
where a ``metric'' $c_{ab}$ is introduced to raise and lower the indices labeling the doubled variables: $(1,2)$ in (\ref{eq:hamiltonian1}) and $(+, -)$ in (\ref{eq:hamiltonian1pm}). For the former $c_{ab} = {\rm diag}(1,-1)$ and for the latter $c_{ab} = {\rm offdiag}(1,1)$ so that $\vecp_a \dot{\vecq}^{\,a} = c^{ab} \vecp_a \dot{\vecq}_b$ (repeated indices are summed) where $c^{ab}$ is the inverse of $c_{ab}$.
Define new Poisson brackets by
\begin{align}
	\{\{ f, g \}\} \equiv  \frac{ \partial f }{ \partial \vecq^{\,a} } \cdot \frac{ \partial g }{ \partial \vecp_a } - \frac{ \partial f }{ \partial \vecp_a } \cdot \frac{ \partial g }{ \partial \vecq^{\,a} }  ,
\label{eq:poisson1}
\end{align}
which can be shown to satisfy Jacobi's identity. Then, Hamilton's equations follow by extremizing the action (\ref{eq:action3}), 
giving
\begin{align}
	\dot{\vecq}_a = \frac{ \partial A }{ \partial \vecp^{\,a} } = \{\{ \vecq_a, A \}\}  , ~ \dot{\vecp}_a = - \frac{ \partial A}{ \partial \vecq^{\,a} } = \{\{ \vecp_a, A \}\} .
\label{eq:hamiltonseq1}
\end{align}
Note the index positions since they are raised and lowered by the metric $c_{ab}$. In the physical limit, (\ref{eq:hamiltonseq1}) becomes Hamilton's equations for a non-conservative system,
\begin{align}
	\dot{\vecq} & = \frac{ \partial H }{ \partial \vecp }  - \bigg[ \frac{ \partial K }{ \partial \vecp_- } \bigg]_{ {\rm p.l.}} \!\!\!\! = \{ \vecq, H \} - \big[ \{\{ \vecq_- , K \}\} \big]_{ {\rm p.l.}  } , \nonumber \\ 
	~ \label{eq:hamiltonseq2} \\
	\dot{\vecp} & = - \frac{ \partial H}{ \partial \vecq } + \bigg[ \frac{ \partial K}{ \partial \vecq_- }  \bigg]_{ {\rm p.l.} } \!\!\!\! = \{ \vecp, H \} - \big[ \{\{ \vecp_-, K \}\} \big]_{ {\rm p.l.} } \nonumber  .
\end{align}

The total time derivative of  the {\it energy function} \cite{Goldstein},
\begin{align}
	h ( \vecq, \dot{\vecq}) = \dot{\vecq} \cdot \frac{ \partial L }{ \partial \dot{\vecq}} - L ,
\label{eq:energyfn2}
\end{align}
follows from the usual manipulations \cite{Goldstein}, which here give
\begin{align}
	\frac{ d h }{ dt } = - \frac{ \partial L }{ \partial t } - \dot{\vecq} \cdot \bigg[ \frac{ d}{dt} \frac{ \partial K }{ \partial \dot{\vecq}_- } - \frac{ \partial K }{ \partial \vecq_- } \bigg] _{\rm p.l.}  .
\label{eq:energyfn3}
\end{align}
The amount of energy entering or leaving the system is determined by $K$ when $\partial L / \partial t=0$ and can be found directly from the new Lagrangian.

~\\
\noindent {\it Example: Viscous drag forces}.  
This new formalism can be used like the standard theory. Consider the following new Lagrangian, given in the ``$\pm$'' variables,
\begin{align}
	\Lambda ( \vec{x}_\pm, \dot{\vec{x}}_\pm) = {} & m \, \dot{\vec{x}}_- \! \cdot  \dot{\vec{x}}_+ - \alpha \,  \vec{x}_- \! \cdot   \dot{\vec{x}}_+ \, | \dot{\vec{x}}_+ | ^{n-1}
\label{eq:draglagrangian1}
\end{align}
where $n=1$ (linear) or $2$ (nonlinear). The first term is the difference of the two kinetic energies ($=m \dot{\vec{x}}_1^{\,2} / 2 - m \dot{\vec{x}}_2^{\,2} / 2$), and the second term is $K$.
The new Lagrangian (\ref{eq:draglagrangian1}) is unique up to terms nonlinear in $\vec{x}_-$ and its time derivatives, which don't contribute to physical forces (see (\ref{eq:shortcut1})).
Using (\ref{eq:shortcut1}), or (\ref{eq:newEL1}) and (\ref{eq:newp1}), gives the equations of motion in the physical limit, $m \ddot{x}^i = - \alpha \, \dot{x}^i |\dot{\vec{x}} \, | ^{n-1}$. For $n=1$ the force is proportional to $-\dot{x}^i$ and for $n=2$ it is proportional to $-\dot{x}^i |\dot{\vec{x}} \, |$.
The former is Stokes' law for the drag force on a spherical object moving slowly through a viscous fluid and the latter is a nonlinear drag force for motions with large Reynolds number \cite{Batchelor}. 
The key point is that these (nonlinear) equations for dissipative motion are derived from a (new) Lagrangian.

To show that the resulting solutions from initial data are consistent with the new Hamilton's principle, it is sufficient to consider slow motions ($n=1$) for which the equations of motion are linear. In the ``$\pm$'' variables the new Euler-Lagrange equations are $m \ddot{x}^i _\pm = - \alpha \dot{x}^i_\pm$.  
The physical limit implies that $\vecx_+$ is determined by the physical initial data,
$\vecx_+ (t_i) = \vecx_i$ and $\dot{\vecx}_+(t_i) = \vecv_i$, while $\vecx_-$ is specified by {\it final} data, $\vecx_- (t_f) = 0 = \dot{\vecx}_-(t_f)$, according to the equality condition. 
Because $\vecx_-$ does not survive the physical limit, prescribing (trivial) data for $\vecx_-$ at the final time is of no physical consequence. The resulting solutions are $\vecx_- (t) = 0$ and $\vecx_+ (t) = \vecx_i + m \vecv_i/\alpha \, [ 1- e^{-\alpha (t-t_i)/m} ]$.
The former automatically imposes the physical limit so that $\vecx_+(t)$ is the physically correct solution.
The new action is stationary for these solutions, as can be shown by direct substitution into (\ref{eq:variation2}).

With $K$ given by the second term of (\ref{eq:draglagrangian1}) it follows from (\ref{eq:energyfn2}) and (\ref{eq:energyfn3}) that $h = m \dot{\vec{x}}^{\,2} / 2$ and $dh / dt = - \alpha | \dot{\vec{x}} \, |^{n+1}$, which is precisely the energy lost per unit time by the object through frictional forces from viscous drag.

~\\
\noindent {\it Example: Coupled harmonic oscillators}.  Return to the first example of a harmonic oscillator $q$ coupled to another oscillator $Q$ to show that the new framework gives the correct physical description for the open dynamics of $q$ itself. Assume initial conditions $q(t_i) = q_i$, $\dot{q}(t_i) = v_i$, $Q(t_i) = Q_{i}$, and $\dot{Q} (t_i) = V_{i}$. The \emph{total} system is closed implying that $K=0$ and the usual action is given by  (\ref{eq:action1}).
Doubling the degrees of freedom, the new action is constructed as in (\ref{eq:action3}) but with $K=0$.
The effective action for the \emph{open} dynamics of the $q$ oscillator subsystem itself is obtained by integrating out the $Q_{\pm}$ variables, which satisfy (\ref{eq:EL2}), $M \ddot{Q}_{\pm} + M \Omega^2 Q_{\pm} = \lambda q_\pm$. Subject to the initial conditions and the equality condition at the final time, the solutions are
\begin{align}
	Q_{+}  (t) = {} & Q^{(h)} (t) + \frac{ \lambda }{ M } \int _{t_i}^{t_f} \!\!\! dt' \, G_{\rm ret} (t-t') q_+ (t') \\
	Q_{-} (t) = {} & \frac{ \lambda }{ M } \int_{t_i}^{t_f} \!\!\! dt' \, G_{\rm adv} (t-t') q_- (t') 
\end{align}
where $Q^{(h)}(t) = Q_{i} \cos \Omega (t-t_i) + V_{i} / \Omega \sin \Omega (t-t_i)$ is the homogeneous solution. The ``$+$'' variable evolves forward in time and satisfies the initial conditions while the ``$-$'' variable evolves backward in time because of the equality condition at the final time. This is a general feature of the ``$\pm$'' variables.

Substituting these solutions into the action yields the effective action,
\begin{align}
	\! S_{\rm eff} [ q_\pm ] = {} &  \int_{t_i}^{t_f} \!\!\! dt \bigg\{ m \big( \dot{q}_+ \dot{q}_- - \omega^2 q_+ q_- \big)  +  \lambda q_-  Q^{(h)} \nonumber \\
		& + \frac{ \lambda^2 }{ M } \int_{t_i}^{t_f} \!\!\! dt' \, q_-(t) G_{\rm ret} (t-t') q_+ (t') \bigg\} .
\label{eq:effaction2}
\end{align}
The factor $q_-(t) q_+(t')$ in the last term is not symmetric in $t \leftrightarrow t'$ and couples to the full retarded Green function as opposed to just its time-symmetric piece as in (\ref{eq:qeom1}). 
Applying (\ref{eq:shortcut1}) to (\ref{eq:effaction2}) gives the equation of motion
\begin{align}
	m \ddot{q} + m \omega^2 q = \!  \frac{ \lambda^2}{ M } \! \int _{t_i}^{t_f} \!\!\!\! dt' \,  G_{\rm ret} (t-t') q (t')  + \lambda Q^{(h)} (t)
\label{eq:qeom2}
\end{align}
in the physical limit.
Now, the Green function in (\ref{eq:qeom2}) is the retarded one, $G_{\rm ret} (t-t') =  \theta (t-t') / \Omega  \sin \Omega (t-t')$, and solutions to (\ref{eq:qeom2}) evolve causally from initial data. 

Generalizing to $N$ oscillators, $Q \to \{Q_n\}_{n=1}^N$, it is straightforward to show that the effective Lagrangian, from $S_{\rm eff} = \int dt \, \Lambda_{\rm eff}$, is
\begin{align}
	\Lambda_{\rm eff} (q_\pm, \dot{q}_\pm) = {} & m \big( \dot{q}_- \dot{q}_+ - \omega^2 q_- q_+ \big) + q_- F(t) \nonumber \\
		& + \int_{t_i}^t dt' \, q_- (t) \gamma(t-t') q_+(t') .
\label{eq:efflagrangian1} 
\end{align}
Here, $F(t) \equiv \sum_{n=1}^N \lambda_n Q_n^{(h)}(t)$ acts like an external force and $\gamma (t-t') \equiv \sum_{n=1}^N \lambda_n^2 / (M_n \Omega_n ) \sin \Omega_n (t-t')$ where a quantity with a subscript $n$ is associated with $Q_n$.
The last two terms in (\ref{eq:efflagrangian1}) 
constitute an effective non-conservative potential, $K_{\rm eff}$, for the open subsystem that is non-local in time and history-dependent. 

From (\ref{eq:energyfn3}), the energy function evolves as
\begin{align}
	\frac{ d h }{ dt } = {} & \dot{q} F(t) + \dot{q} \int_{t_i}^t dt' \, \gamma(t-t') q(t')
\label{eq:dhdtosc1}
\end{align}
where $h= m (\dot{q}^2 + \omega^2 q^2)/2$ is the energy of the oscillator from (\ref{eq:energyfn2}). 
To see a familiar dissipation, choose trivial initial data for the $\{ Q_n \}$ so that $F(t) = 0$ and take each $M_n$ to be a constant, $M$. The coupling strengths $\{\lambda_n\}$ are arbitrary so let $\lambda_n = \lambda \Omega_n$ for  $\lambda$ constant. Then, $\gamma (t-t') = ( \lambda^2 / M) d/dt' \sum_{n=1}^N \cos \Omega_n (t-t')$. If $N$ is so large that $q$ essentially couples to a continuum of oscillators then the summation becomes integration over $\cos \Omega (t-t')$, which is a Dirac delta distribution (local in time). With these considerations, the frequency $\omega^2$ is renormalized to $\omega_{\rm ren}^2 = \omega^2 - \delta(0) \lambda^2/(mM)$ and (\ref{eq:dhdtosc1}) becomes $dh / dt = - \gamma_0 \dot{q}^2 (t)$
for $\gamma_0 = \lambda^2/(2M)$, 
which is the power lost by a damped, simple harmonic oscillator.

 ~\\
\noindent {\it Concluding remarks}. 
The main results of this paper include the construction of a variational principle for initial value problems and the formulation of Lagrangians and Hamiltonians for general non-conservative systems.
The key aspects of this classical mechanics are the formal doubling of variables and the $K$ function describing non-conservative forces and interactions. 
For demonstrative purposes I have focused on discrete mechanical systems but the formalism is equally applicable to continuum systems like field theories (see \cite{GalleyLeibovich:PRD86} for a non-trivial application) and elastic media. 

An open system, which can exchange energy by interaction with some other set of variables, will have a non-vanishing $K$. Generally, there are two scenarios when this happens: 1) When the underlying variables that cause the non-conservative (e.g., dissipative) forces are neither given nor modeled so that $K$ must be \emph{prescribed}; and 2) When all the degrees of freedom of a total (i.e., closed) system are given or modeled, and a suitable subset of those variables are integrated out leaving the remaining open subsystem described by a \emph{derived} $K$. The first scenario encompasses the viscous drag example where $K$ is prescribed so that the resulting drag force is the desired one. The second scenario includes the coupled oscillators example where $K$ is derived for the open subsystem $q(t)$ by integrating out the $\{Q_n\}$ (see discussion after (\ref{eq:efflagrangian1})). 

The formalism developed here can be canonically quantized by replacing the new Poisson brackets in (\ref{eq:poisson1}) by commutators. Similarly, one can implement a path integral quantization using the new action (\ref{eq:action3}). The results of this paper thus provide a foundation for quantizing non-conservative systems where $K$ is \emph{prescribed}. For open quantum systems where $K$ is \emph{derived} one often uses the so-called ``in-in'' quantum theory \cite{Schwinger:JMathPhys2} or the closely related Feynman-Vernon formalism \cite{FeynmanVernon:AnnPhys24}. Such studies apply to cases where the environment is given or modeled. Quantization where $K$ is \emph{prescribed} thus generalizes the usual in-in formalism to systems like the viscous drag example.

The new formulation of non-conservative systems constructed here may be useful for any method or technique that normally uses, or could benefit from using, Lagrangians and Hamiltonians. 
These might include: developing partition functions for non-conservative statistical systems (see also \cite{TuckermanMundyMartyna}), studying the phase space structure of nonlinear dissipative dynamical systems, and developing variational numerical integrators for systems with physical dissipation, 
 among others.
Also, the appearance of a metric in (\ref{eq:hamiltonian2}), the hint of ``covariance'' in (\ref{eq:hamiltonian1}) and (\ref{eq:hamiltonian1pm}), and the use of doubled variables suggest additional structure for the symplectic manifold \cite{ArnoldCM}.
In \cite{GoldbergerRothstein:PRD73_2}, extra physical degrees of freedom are introduced in a Lagrangian to parameterize absorptive processes within the paradigm of effective field theory (EFT) (see also \cite{Porto:PRD77, LopezNacir:2011kk} for recent applications). That work, in combination with results presented here, may provide a powerful tool for studying dissipative systems that also satisfy the underlying assumptions of EFT.


I thank Y.\,Chen, C.\,Cutler, K.\,Hawbaker, A.\,Leibovich, H.\,Miao, E.\,Poisson, I.\,Rothstein, G.\,Sch\"afer, L.\,Stein, A.\,Tolley, M.\,Vallisneri, and especially A.\,Zengino\u{g}lu for discussions and comments of previous drafts. This work was supported in part by an appointment to the NASA Postdoctoral Program at the Jet Propulsion Laboratory administered by the Oak Ridge Associated Universities through a contract with NASA.  Copyright 2012. All rights reserved.

\bibliographystyle{physrev}
\bibliography{gw_bib}

\setlength{\parskip}{1em}

\end{document}